\documentclass[a4paper]{jpconf}
\usepackage{graphicx}
\usepackage{iopams}
\usepackage[square,sort&compress,numbers]{natbib}
\def\schpt{S\raise0.4ex\hbox{$\chi$}PT}
\def\chpt{\raise0.4ex\hbox{$\chi$}PT}

\begin{document}
\title{Electromagnetic effects on the light hadron spectrum}

\author{S Basak$^1$, A Bazavov$^2$, C Bernard$^3$, C DeTar$^4$, E Freeland$^5$,  J Foley$^4$,
Steven Gottlieb$^6$,
U M Heller$^7$, J Komijani$^3$, J Laiho$^8$, L Levkova$^4$, R~Li$^6$, 
J Osborn$^9$, R~L~Sugar$^{10}$,
A Torok$^6$,  
D Toussaint$^{11}$, R~S~Van~de~Water$^{12}$ and 
R Zhou$^{12}$ [MILC Collaboration]}

\address{$^1$ NISER, Bhubaneswar, Orissa 751005, India}
\address{$^2$ Department of Physics and Astronomy,
University of Iowa, Iowa City, IA 52240, USA}
\address{$^3$ Department of Physics, Washington University, St. Louis, MO 63130, USA}
\address{$^4$ Department of Physics and Astronomy, University of Utah, Salt Lake City, UT 84112, USA}
\address{$^5$ Liberal Arts Department, School of the Art Institute of Chicago, Chicago, IL, USA}
\address{$^6$ Department of Physics, Indiana University, Bloomington, IN 47405, USA}
\address{$^7$ American Physical Society, One Research Road, Ridge, NY 11961, USA}
\address{$^8$ Department of Physics, Syracuse University, Syracuse, NY  13244, USA}
\address{$^9$ ALCF, Argonne National Laboratory, Argonne, IL 60439, USA}
\address{$^{10}$ Physics Department, University of California, Santa Barbara, CA 93106, USA}
\address{$^{11}$ Physics Department, University of Arizona, Tucson, AZ 85721, USA}
\address{$^{12}$ Theoretical Physics Department, Fermilab, Batavia, IL 60510, USA}

\ead{sg@indiana.edu}

\begin{abstract}
For some time, the MILC Collaboration has been studying electromagnetic 
effects on light mesons.  These calculations use fully dynamical QCD, but
only quenched photons, which suffices to NLO in \chpt.  That is, the sea quarks are electrically neutral,
while the valence quarks carry charge.  For the photons we use
the non-compact formalism.  We have new results with lattice spacing
as small as 0.045 fm and a large range of volumes.  
We consider how well chiral perturbation theory describes these results and
the implications for light quark masses.

\end{abstract}

\section{Introduction}
The MILC collaboration has been studying electromagnetic and isospin-violating 
effects in pions and kaons for a number of years
\cite{Basak:2008na,Torok:2010zz,Basak:2012zx,Basak:2013iw}.  These effects are
crucial for determining the three light-quark masses.  The $u$, $d$, and $s$
quark masses are both fundamental parameters of the standard model of
elementary particle physics and important for phenomenology.  The uncertainty
in the size of the electromagnetic contributions to the pion and kaon
masses is a major part of the error on each of the masses and the
greatest source of error on the $m_u/m_d$ mass ratio \cite{Bazavov:2009bb}.

The electromagnetic error in $m_u/m_d$ depends on the error in our estimate of 
$(M^2_{K^+} - M^2_{K^0})^\gamma$ where $\gamma$ indicates the electromagnetic
contribution to the difference in the squares of the charged and neutral
kaon masses.  (There is also a difference coming from the quark masses.)  
In 1960, Dashen \cite{Dashen:1969eg}
showed that at leading order electromagnetic mass splittings for
the mesons are mass independent, {\it i.e.}, 
$(M^2_{K^+} - M^2_{K^0})^\gamma = (M^2_{\pi^+} - M^2_{\pi^0})^\gamma$.
We can parameterize the higher order effects via a parameter $\epsilon$
defined by
$(M^2_{K^+} - M^2_{K^0})^\gamma = 
(1+\epsilon)(M^2_{\pi^+} - M^2_{\pi^0})^{\rm exp}$.
This definition of $\epsilon$, which uses the experimental mass difference
on the RHS, is slightly different from what we have used before, in which
$(M^2_{\pi^+} - M^2_{\pi^0})^\gamma$ appears on the RHS.

\begin{table}[tbh]
\caption{Parameters of the (2+1)-flavor asqtad ensembles used in this study. 
Volumes marked with $*$ are currently used in the finite volume studies, but not
in the full analysis. The quark masses $m'_l$ and $m'_s$ are the light 
and strange dynamical masses used in the runs.  
The  number of configurations listed  as `132+52' for the 
$a\!\approx\!0.12\:$fm, $48^3\times 64$ ensemble gives values for
two independent streams, the first in single precision, 
and the second in double.  
At the moment, we treat them as separate data, and do not average the results. 
\label{tab:ensembles} }
\begin{center}
\begin{small}
\begin{tabular}{lllllll}
\br
$\approx a$[fm]& Volume
& $\beta$
& $m'_l/m'_s$& \# configs.  &$L$ (fm) & $m_\pi L$ \\
\mr
0.12 & $12^3\times64^*$ & 6.76& 0.01/0.05&  1000  & 1.4 & 2.7    \\
     & $16^3\times64^*$ & 6.76& 0.01/0.05&  1303  & 1.8 & 3.6   \\
     & $20^3\times64$ & 6.76& 0.01/0.05&  2254 & 2.3 &  4.5  \\
      & $28^3\times64$ & 6.76& 0.01/0.05& \phantom{2}274 & 3.2& 6.3   \\
      & $40^3\times64^*$ & 6.76& 0.01/0.05& \phantom{2}115 & 4.6& 9.0   \\
      & $48^3\times64^*$ & 6.76& 0.01/0.05& 132+52 & 5.4& 10.8   \\
      & $20^3\times64$ & 6.76& 0.007/0.05& 1261 & 2.3& 3.8   \\
      & $24^3\times64$ & 6.76& 0.005/0.05& 2099 & 2.7& 3.8   \\
\mr
0.09  & $28^3\times96$ &7.09& 0.0062/0.031& 1930 & 2.3&  4.1  \\
      & $40^3\times96$ & 7.08& 0.0031/0.031& 1015 & 3.3& 4.2   \\
\hline
0.06  & $48^3\times144$ & 7.47& 0.0036/0.018& \phantom{2}670 & 2.8& 4.5   \\
  & $56^3\times144$ & 7.465& 0.0025/0.018& \phantom{2}433 & 3.3& 4.4   \\
  & $64^3\times144$ & 7.46& 0.0018/0.018& \phantom{2}826 & 3.7& 4.3   \\
\mr
0.045  & $64^3\times192$ & 7.46& 0.0028/0.014& \phantom{2}861 & 2.8& 4.6   \\
\br
\end{tabular}
\end{small}
\vspace{-8mm}
\end{center}
\end{table}

Our calculation is done with dynamical QCD but with quenched, noncompact
QED.  The analysis  of Bijnens and Danielsson \cite{Bijnens:2006mk}
shows that quenched QED is sufficient for a controlled calculation of
$\epsilon$ at NLO in $SU(3)$ chiral perturbation theory.  

In table \ref{tab:ensembles}, we list the ensembles of gauge configurations 
used for the results reported here.  The two larger volume 
$a\!\approx\!0.06\:$fm and the $a\!\approx\!0.045\:$fm ensembles were not
included in our results presented at Lattice 2014 \cite{Basak:2014vca}.
The $40^3\times64$ and $48^3\times 64$ ensembles with $\beta=6.76$ were 
generated after Lattice 2014 and first discussed at this conference, but they
are included in Ref.~\cite{Basak:2014vca}.

\section{Finite volume effects}
\begin{figure}
\vspace{-5mm}
\begin{center}\includegraphics[width=0.53\textwidth]{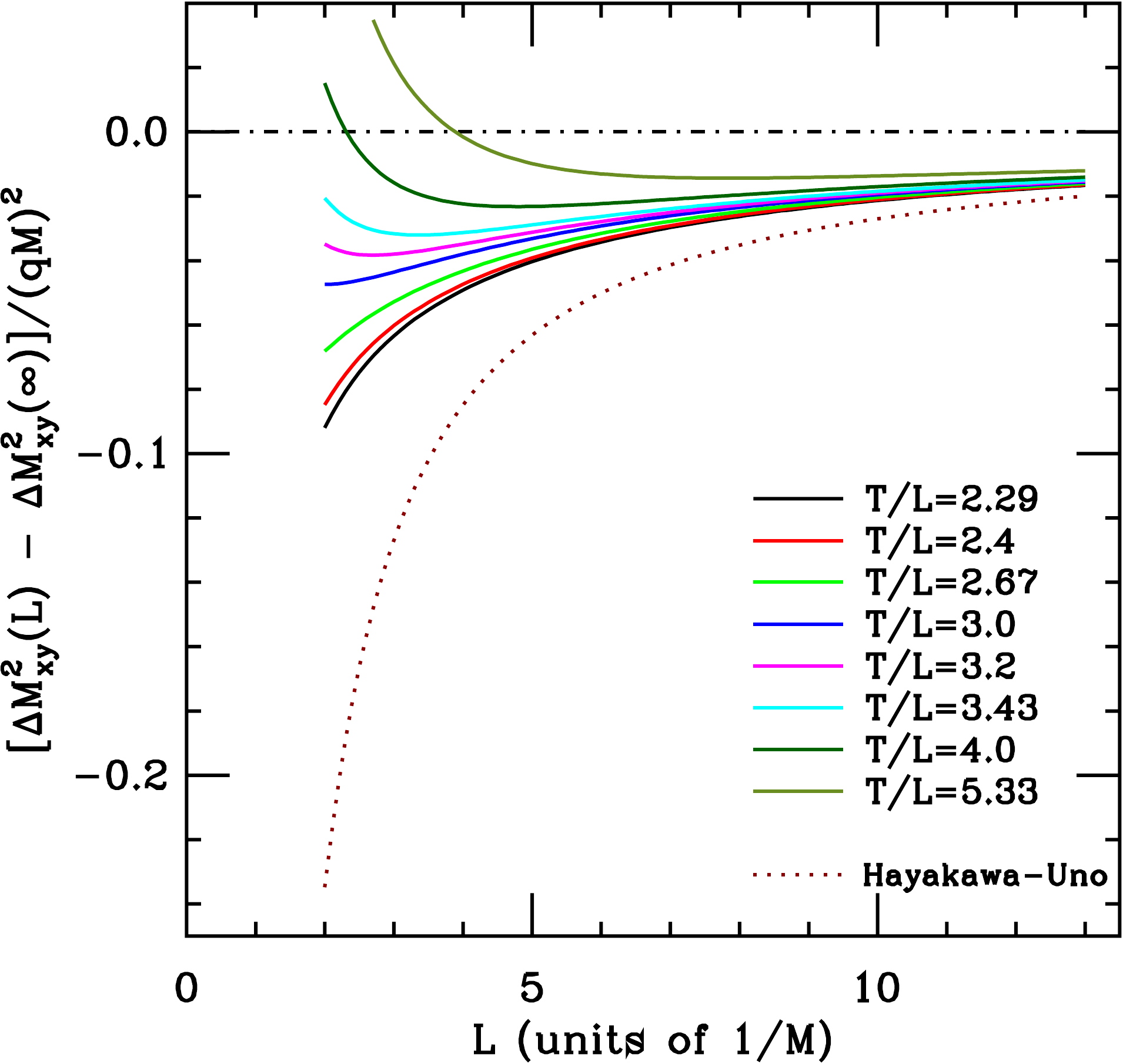}\end{center}
\vspace{-5mm}
\caption{\label{fig:MILC-HU} 
Comparison of electromagnetic finite volume (FV) effects in our scheme
with that in Ref.~\cite{Hayakawa:2008an}. 
$q$ and $M$ are the charge and mass of the meson, respectively.
The values of $T/L$ correspond to aspect ratios used in our ensembles, with
4.0 and 5.33 corresponding to the two smallest volumes in our FV
study.  Our FV effects are a factor of 2--3 smaller in most
of the relevant range.
}
\vspace{-0.15in}
\end{figure}

Because the photon is massless, finite volume effects are an important issue
and a source of uncertainty in our earlier results.  For $a\!\approx\!0.12\:$fm,
we had volumes of $20^3\times 64$ and $28^3\times64$ and found small finite
volume effects compared to what was seen in Ref.~\cite{Portelli:2012pn}.
Electromagnetic finite volume effects have been calculated by 
Hayakawa and Uno~\cite{Hayakawa:2008an}
using chiral perturbation theory.  They found rather large effects; however,
there is not a unique choice of how to treat finite-volume
zero modes for the $U(1)$ field in the noncompact
formalism.  For the temporal modes $A_0$ in Coulomb gauge, 
all modes for 3-momentum $\vec 
k=0$
must be dropped.  That is, $A_0(\vec 0, k_0) = 0$ for all $k_0$.  For the
spatial modes, the action density is 
$[(\partial_0 A_i)^2 + (\partial_j A_i)^2)]/2$, so the only mode that
must be dropped is that with $(\vec 
k, k_0) = (\vec %
0, 0)$.
Hayakawa and Uno dropped all $A_i$ modes with $\vec 
k=0$.  However, we have
only dropped modes with both  $\vec 
k=0$ and $k_0=0$.  In our case, the
finite size effects are smaller, although the magnitude of the effect does
depend on $T/L$ where $T$ ($L$) is the temporal (spatial) extent of the lattice.
Figure \ref{fig:MILC-HU} compares the finite volume effects for the two
choices.

\begin{figure}[bth]
\vspace{-5mm}
\begin{center}\includegraphics[width=0.53\textwidth]{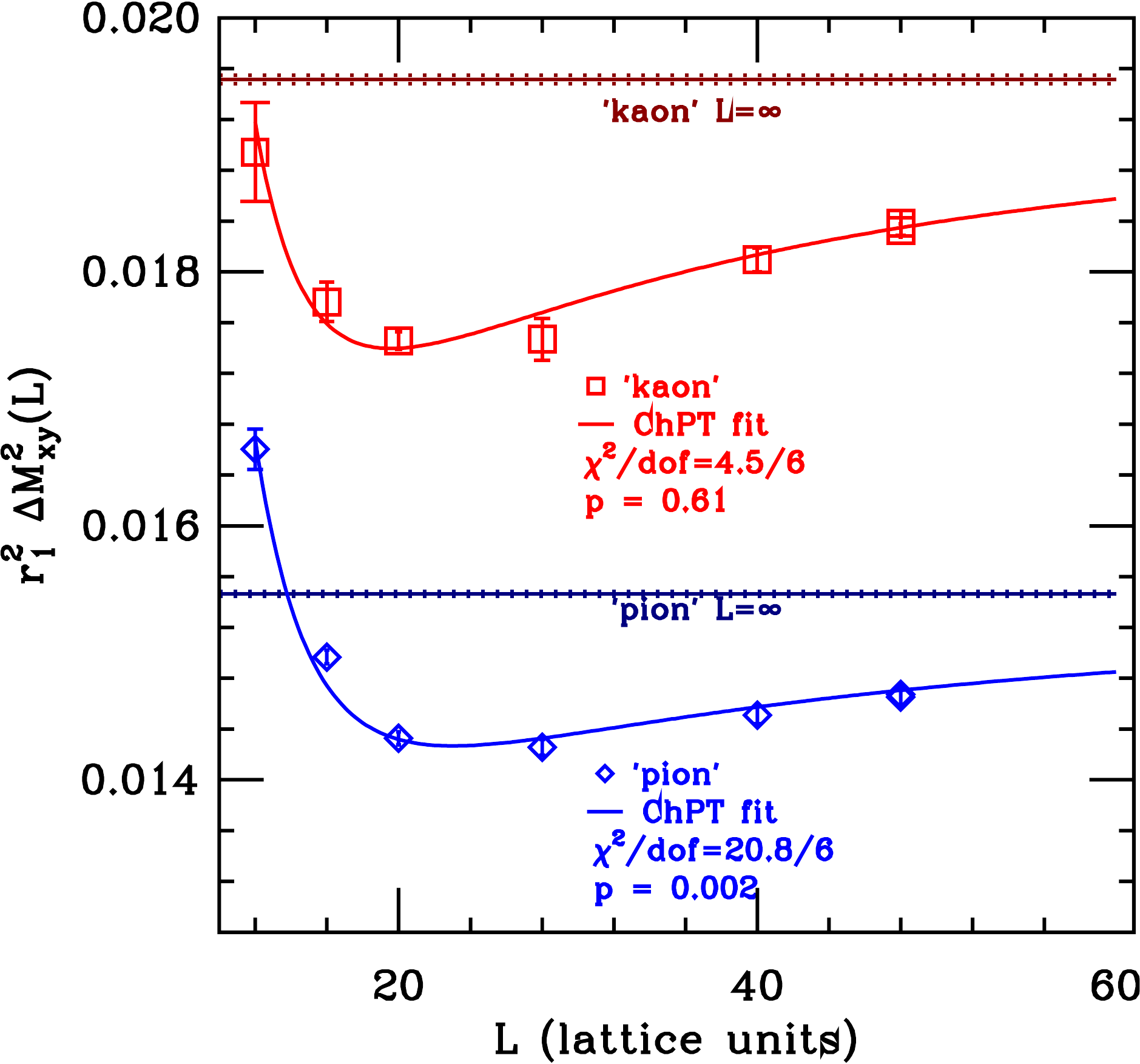}\end{center}
\vspace{-5mm}
\caption{\label{fig:FV} Finite volume effects at $a\approx0.12$ fm  and $am'_l=0.01, am'_s=0.05$ as a function of spatial lattice length $L$ for
two different meson masses: a unitary `pion' (blue) with degenerate valence masses $m_x=m_y=m'_l$, and a `kaon' (red) with valence masses
$m_x=m'_l$ and $am_y=0.04$, close to the physical strange quark mass.  The fit lines are to the FV form from \schpt, and have one free parameter each, the infinite volume value (shown by horizontal solid lines with dotted lines for errors).
}
\vspace{-0.15in}
\end{figure}

In order to test the prediction, we added four new volumes this year 
with $L=12$, 16, 40, and 48.  In Fig.~\ref{fig:FV}, we show our results for six
values of $L$.  
We plot electromagnetic splittings of squared masses 
$\Delta M^2_{xy}\equiv M^2_{xy}- M^2_{x'y'}$, where the first meson is
constructed from valence quarks  $x$ and $y$ with charges
$q_x$, $q_y$ and masses $m_x$, $m_y$, and the second meson is made
from quarks $x'$ and $y'$  with the same masses but with
the quark charges set to zero.  To compare results from different
ensembles, we multiply the squared-mass difference by $r_1^2$ where
$r_1$ is a length scale determined from the static potential.
Each curve is a fit with a single parameter, the
infinite volume limit.  In other words, the shape of the curve, but not 
its height, is determined by the theory.
The horizontal
lines show the infinite volume limit for two mass combinations labeled
`pion' and `kaon.'  It is now clear why we saw such a small difference between
$L=20$ and 28.  The slope of each curve is low in that region.  It is also
worth noting that for the `pion,' the sign of the effect changes within
the range of our new calculation.  With our improved understanding of the
finite volume effects, we can more accurately adjust our results to
the infinite volume limit with a
smaller residual error.  This reduces our
errors on $\epsilon$ and $m_u/m_d$, as we show later.

\section{Chiral fit and extrapolation}
We have calculated meson masses for a large number of valence-quark mass 
and charge combinations.  On some ensembles, we include charges larger than
the physical values.  On others, we calculated all combinations of
charges $\pm 2/3e$, $\pm 1/3e$, and 0.
In the first step of the analysis, values are corrected for finite-volume
effects.   The corrections are 7--10\% for pions (mesons with two light
quarks) and 10--18\% for kaons (mesons with one light quark and one
whose mass is near the strange quark mass).  
Because of the high level of
correlation among the many meson masses, we thin the data set and
sometimes have to consider uncorrelated fits.  
A typical fit might include 150 points.

\begin{figure}[h]
\begin{minipage}{75mm}
\includegraphics[width=75mm]{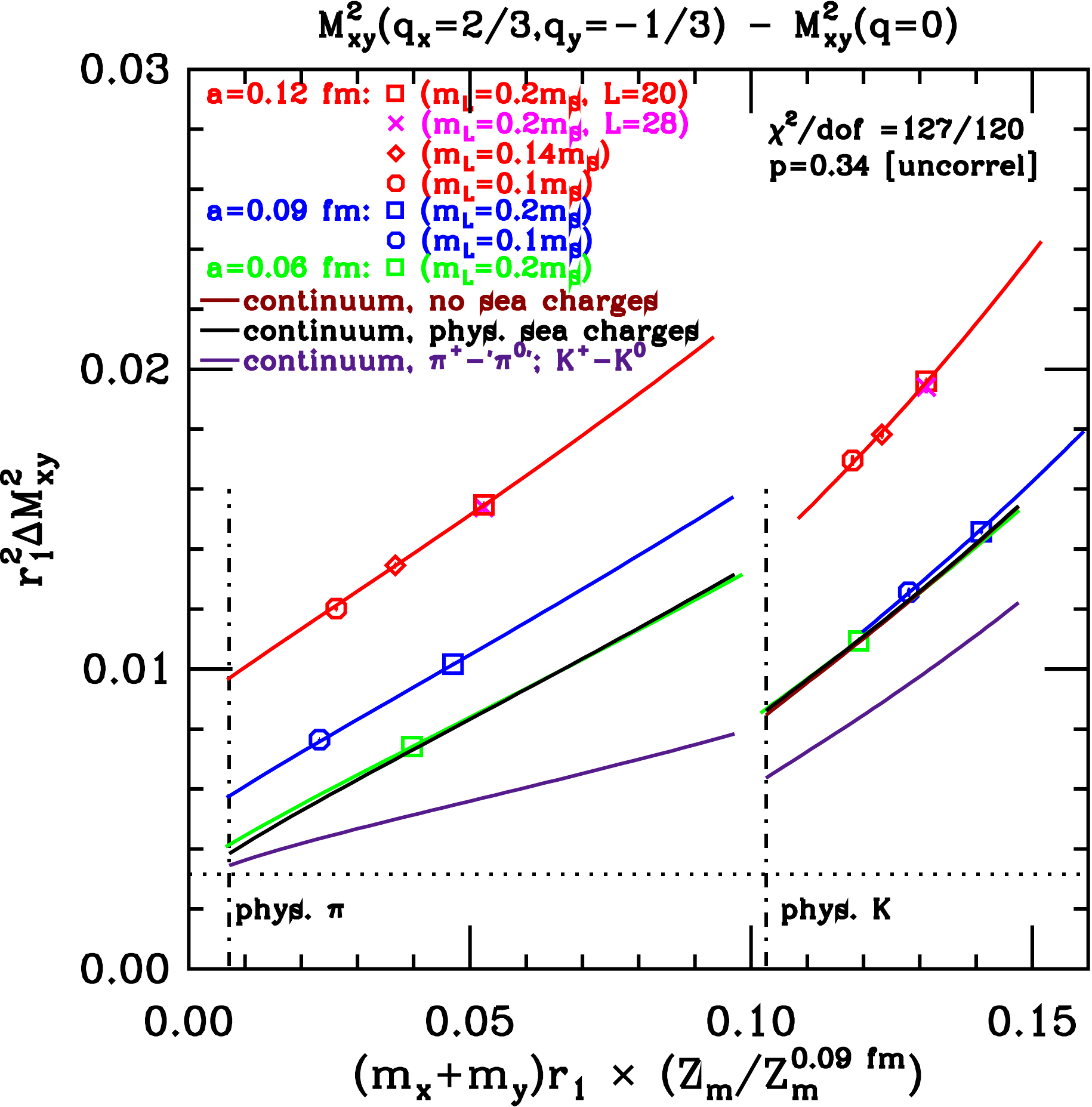}
\end{minipage}\hspace{9mm}%
\begin{minipage}{75mm}
\includegraphics[width=75mm]{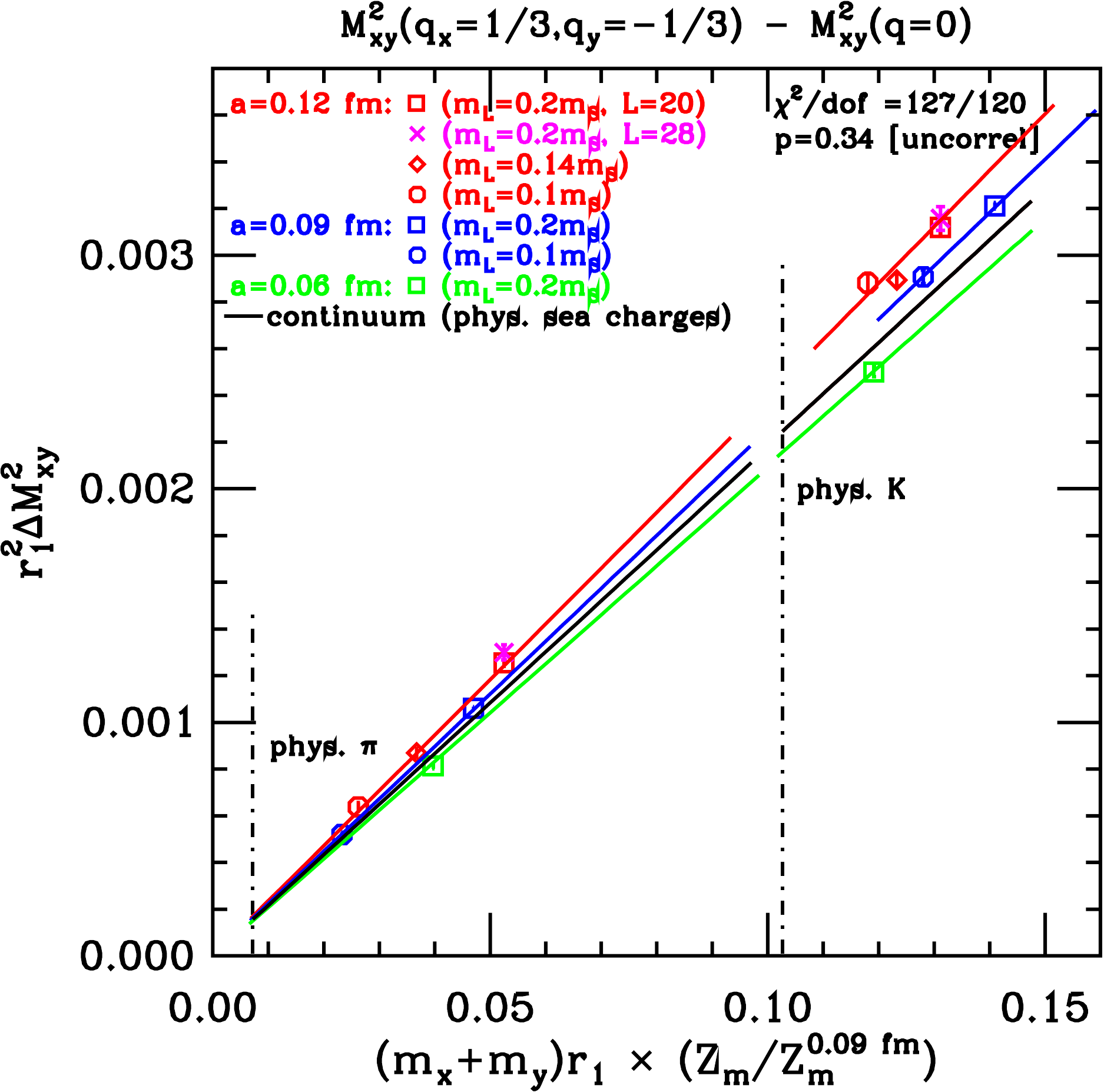}
\end{minipage}
\caption{\label{fig:chiral}
Central fit to  the EM splitting $\Delta M_{xy}^2$ {\it vs}.\ the
sum of the valence-quark masses. 
Only a small subset of the partially quenched data set included in the fit 
is shown:  the points for $a\!\approx\!0.09$  fm and 
$\!\approx\!0.06$ fm, as well as the `pion' ($x$-axis value $<$ 0.07) points for $a\!\approx0.12$ fm, have unitary values of the valence masses, while the
`kaon' ($x$-axis value $>$ 0.1) points for $a\!\approx0.12$ fm have $m_x=m'_l$ but $m_y=0.8m'_s$, which is closer to the physical strange mass than
$m'_s$ itself. 
The data have been corrected for finite volume effects using NLO \schpt.
The red, blue, and green curves correspond to the three  lattice spacings.
The black and purple curves are extrapolations, see text.
}
\end{figure}

Once the chiral fit for uncharged sea quarks is determined, we can
set valence and sea quark masses equal, set $m_s$ to its correct value and
take the continuum limit.  We then
set the sea quark charges to their physical value using the NLO chiral
logs. 
The last adjustment is very small for the kaon
and vanishes identically for the pion.  Figure~\ref{fig:chiral} shows a small
subset of the data for the uncorrelated
fit that determines our central value.  Note how much smaller 
the electromagnetic effects are for the neutral mesons (right).

There is a complication involving the neutral pion.  The physical pion 
propagator involves disconnected diagrams in which the quark and anti-quark
annihilate.  These contributions are difficult to calculate accurately 
because they are noisy.  We drop the disconected
diagrams and use the RMS average mass of $u\bar u$ and $d\bar d$ mesons.  We
denote this state as ``$\pi^0$.''
However, electromagnetic contributions to 
neutral mesons vanish in the chiral limit, so both  the true 
$(M^2_{\pi^0})^\gamma$ and our  $(M^2_{\rm{``}\pi^0\rm{"}})^\gamma$ are
small.  

\begin{figure}
\begin{center}\includegraphics[width=0.53\textwidth]
{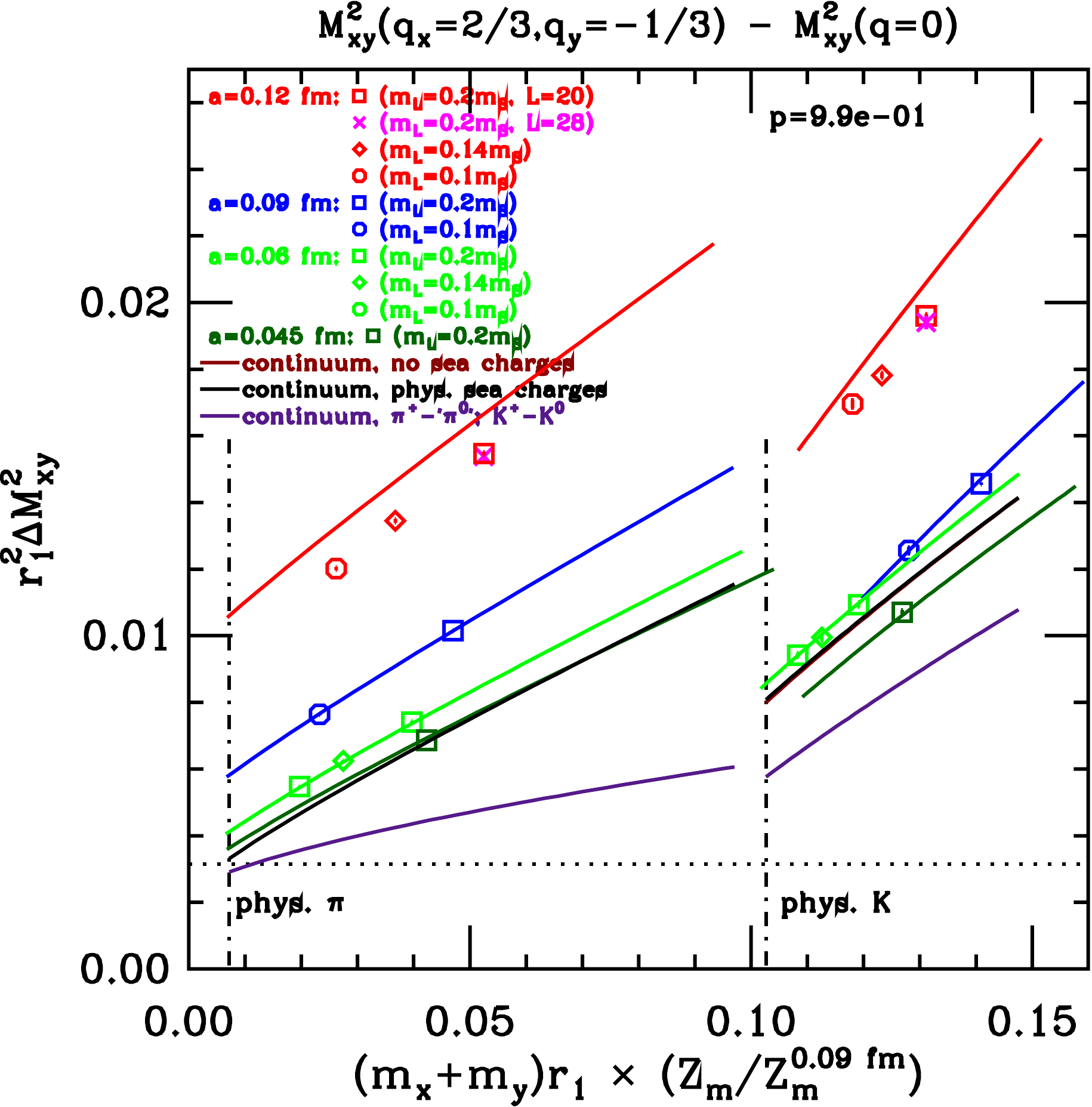}\end{center}
\vspace{-5mm}
\caption{\label{fig:noa0.12} 
Similar to Fig.~\ref{fig:chiral} except that  $a\!\approx\!0.12\:$fm ensembles
are not included in the fit and the correlations are included.  Only a 
subset of the fitted data is shown.
}
\end{figure}

In the last stage of the analysis, we subtract our results for the ``$\pi^0$'' 
and $K^0$ from the corresponding charged meson to arrive at the physical
electromagnetic mass differences.  This is shown as the purple lines on the
LHS of Fig.~\ref{fig:chiral}.  The vertical lines represent the appropriate
sum of quark masses for the charge-averaged pion and kaon masses.  
The horizontal line is the experimental value of the pion splitting.  Thus,
the distance between the horizontal line and the
intersection of the purple line with the vertical,
dashed-dotted physical pion line is an indication of the size of the
systematic error.
Looking at the ratio of the experimental result for the pion splitting to
our kaon splitting, we get $\epsilon= 1.02(4)$.  
Alternatively, we may use our result for the electromagnetic
pion splitting, and we find $\epsilon= 0.84(5)$.  
We have recently added the newer ensembles with
$a\!\approx\!0.06\:$ and 0.045 fm to our analysis.  In this case, we
drop the $a\!\approx\!0.12\:$fm ensembles and can get good correlated 
fits when the data is thinned.  One such fit is shown in Fig.~\ref{fig:noa0.12}.
In this case, we see that our result for the pion splitting is a bit low.
This results in $\epsilon= 0.97(5)$.  If we take the experimental value
of the pion splitting, our result is $\epsilon= 0.83(4)$.

\section{Results and future plans}
Our current (preliminary) result for $\epsilon$ does not include the fit
shown in Fig.~\ref{fig:noa0.12}, but it does include other variations on the 
fits shown in Fig.~\ref{fig:chiral} in which different subsets of our
data are included.  Our result is \cite{Basak:2014vca}
\begin{equation}
\epsilon = 0.84(5)_{\rm stat}(18)_{a^2}(6)_{\rm FV}\;.
\vspace{-1.5mm}
\end{equation}
The first error is statistical, the second from variations in the continuum
extropolation, and the third is our (hopefully conservative)
estimate of the residual finite volume
error that may remain after our correction based on the NLO formula.
Previously, we have determined $m_u/m_d$ by using our pseudoscalar mass results
on our asqtad ensembles.  However, we have recently been generating 
highly improved staggered quark (HISQ) 
ensembles for which the chiral extrapolation is much better controlled.
Using our new value of $\epsilon$ in Eq.~(1) with the HISQ light meson 
masses \cite{Bazavov:2014wgs} gives a preliminary value for the ratio
\begin{equation}
m_u/m_d = 0.4482 (48)_{\rm stat} ({}^{+\phantom{0}21}_{-115})_{a^2} (1)_{\rm FV_{QCD}} (165)_{\rm EM},
\vspace{-1.5mm}
\end{equation}
where here ``EM'' denotes all errors from electromagnetism, while 
``FV$_{\rm QCD}$'' refers to finite-volume effects in the pure QCD calculation.
The electromagnetic error has been reduced by more than a factor of two 
from our previous result \cite{Bazavov:2009bb}.

We plan on several future improvements.  First, we will continue to analyze
the ensembles that were not in our prior work.  We will also repeat the
analysis of quenched electromagnetic effects
on the HISQ ensembles.  The HISQ ensembles have 
several advantages.  They should have smaller discretiziation errors, and the
chiral extrapolation errors should be much smaller, as we have ensembles
tuned to the physical light quark mass.  Finally, the finite volume errors
should be reduced as the HISQ lattices are larger than those for asqtad.
We are also working on a fully dynamical $SU(3) \times U(1)$ code 
to allow us to include charged sea-quark mass effects
\cite{Zhou:2014gga} .  This will make possible controlled calculations
of many additional quantities.  We have also calculated electromagnetic
effects on the baryon spectrum, but did not have time to report on that.

\ack
The spectrum running was done on computers at the National Center for
Supercomputing Applications, Indiana University, the Texas Advanced
Computing Center (TACC), and the National Institute for Computational
Science (NICS). Configurations were generated with resources provided by the
USQCD
Collaboration, the Argonne Leadership Computing Facility, and the National Energy Research Scientific
Computing Center, which are funded by the Office of Science of the U.S.
Department of Energy; and with resources provided by the National Center for Atmospheric
Research,  NICS, the Pittsburgh Supercomputer
Center, the San Diego Supercomputer Center, and TACC,
which are funded through the National Science Foundation's XSEDE Program. This work
was supported in part by the U.S. DOE   
and the NSF. 

\bibliography{embib.bib}
\end{document}